\documentclass[reprint,english,aps,pre,floatfix,showpacs]{revtex4-1}

\usepackage{amsmath}
\usepackage{graphicx}
\usepackage{amssymb}
\usepackage{babel}
\usepackage{color}
\usepackage{bm}
\usepackage{subfigure}

\renewcommand{\vec}[1]{\boldsymbol{#1} }
\makeatletter
\renewcommand{\thesubfigure}{\alph{subfigure}}   
\renewcommand{\@thesubfigure}{\thesubfigure)\hskip\subfiglabelskip}
\makeatother

\begin{document}

\title{Order-disorder transition in swirled granular disks}

\author{Philip Krinninger}
\author{Andreas Fischer}
\author{Andrea Fortini}
\email{andrea.fortini@uni-bayreuth.de}
\affiliation{Theoretische Physik II, Physikalisches Institut, Universit\"at Bayreuth, Universit\"atsstra{\ss}e 30, D-95447 Bayreuth,
Germany}
\pacs{45.70.-n, 64.70.D-, 05.70.Ln}

\begin{abstract}
We study  the order-disorder transition of horizontally swirled dry and wet granular disks  by means of computer simulations.  Our systematic investigation of the local order formation as a function of amplitude and period of the external driving force shows that a large cluster of hexagonally ordered particles forms for both dry and wet granular particles at intermediate driving energies.
Disordered states are found  at small and large driving energies.
Wet granular particles reach a higher degree of local hexagonal order, with respect to the dry case. For both cases we report a qualitative phase diagram showing the amount of local order at different state points. Furthermore we find that the transition from hexagonal order to a disordered state is characterised by the appearance of particles with square local order.
\end{abstract}

\maketitle

\section{Introduction}

The segregation of grains with different sizes is of importance in many industries. In the last 30 years the segregation problem attracted much attention due to the complexity of the Brazil Nut Effect~\cite{Rosato:1987dv}, i.e. the rise of large intruder particles in vertically oscillated granulates. 
Different phenomena contribute to this effect~\cite{Knight:1993bg,Cooke:1996gg,Poschel,Kudrolli:2004kr,schroter:2006cc,Majid:2009jl} and their characterisation is complex due to the large number of parameters controlling the dynamics of the system. Among these parameters, the gravitational force and the height of the granular bed play an important role.

On the other hand, already the gold miners of the 1899 Klondike river gold rush knew that by applying a swirling motion to a pan of sand the desired gold nuggets were slowly exposed~\cite{lagal}. 
The simple horizontal movement  can be used to obtain size segregation without the influence of  the gravitational force.  The granular bed's height is also not important when a monolayer of particles is studied.
The occurrence of segregation of differently sized grains under 'swirling' was studied experimentally~\cite{Aumaitre:2001fd,Schnautz:2005bp}, and later the dynamics of a single intruder was analysed by~\citet{Chung:2008jz}.

The complex segregation behaviour lies on top of a rich dynamical behaviour of single sized particles that has been little studied in this system. Work has been done on 
the dynamical motion of the granular matter~
\cite{Scherer:1996hl,Feltrup:2009gm}, as well as the formation of solid-like clusters ~\cite{Kotter:1999iz,Scherer:2000jf}.
Recently,~\citet{May:2013tf} investigated in more detail the formation of solid clusters and found that the melting of the solid clusters occurs first at their surface. 
 
The formation of compact ordered structures can have huge impact on the diffusion of the particles and the segregation of particles of different sizes.
In this paper we present a systematic simulation study of the order-disorder transition that is induced by the horizontal swirling of granular disks.
Within the correct envelope of amplitudes and oscillation periods we find a transition from  disordered fluid-like granular clusters to hexagonally ordered clusters in both dry and wet granular matter. 
At high driving energies  the transition to disordered structure occur due to the strong compression of the granular clusters at the container wall. We find that this transition is characterised by a predominance of particles with local square order. 

This paper is organised as follows. In Sec.~\ref{sec:CM} we briefly introduce the simulation technique and the model used for the description of the granular particles, while results for both dry and wet granular matter are given in Sec.~\ref{res}. In Sec.~\ref{conc} we draw our conclusions, while in appendix~\ref{sec:bondorder} we describe the application of the $q_{6}$ order parameter to our two dimensional system.

\section{Model and Method}

\label{sec:CM}

We study a system of  $N$ monodispersed disks with diameter $\sigma$  and equal mass $m$ in a circular container of radius $R/\sigma=12.25$. The numerical value of the container radius is based on the experiments of~\citet{May:2013tf}.

The pair interaction between particles is modelled by an Hertzian pair-contact collision model. For each contact between two pairs of particles  at positions $\vec  r_i$ and $\vec r_j$, with   velocities $\vec v_i$ and $\vec v_j$, and angular velocities $\vec \omega_i$ and $\vec \omega_j$ we define a contact plane.
\begin{figure}[htbp]\centering
	\includegraphics[width=6cm]{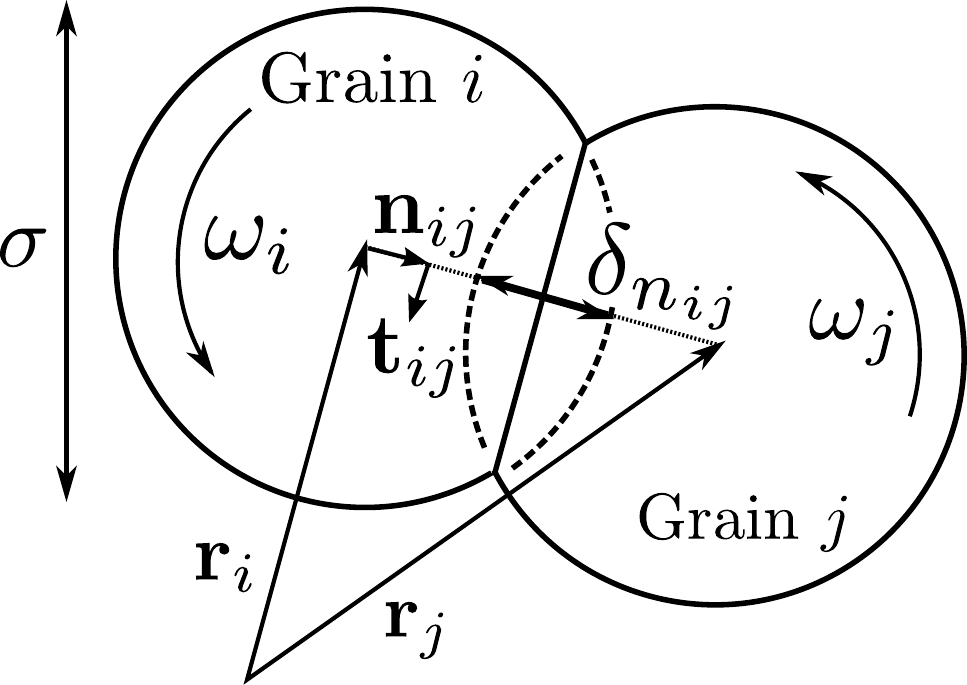}
	\caption{Sketch of the pair contact model in the contact plane defined by the vectors $\vec n_{ij}$ and  $\vec v_{ij}$. }
	\label{force}
\end{figure}
The two vectors that generate the contact plane are the  normal unit vector $\vec n_{ij}=\frac{\vec  r_i-\vec  r_j}{|\vec  r_i-\vec  r_j|}$, and the relative velocity $\vec v_{ij}= \vec v_i -\vec v_j$. 
Figure~\ref{force} shows a sketch of the collision model projected in the collision plane: the total force  is decomposed in the normal direction $\vec n_{ij}$ and the tangential directions $\vec t_{ij}=\frac{\vec v_{t_{ij}}}{|\vec v_{t_{ij}}|}$, where the relative velocities in the normal and tangential directions are
\begin{eqnarray}
\vec v_{n_{ij}}&=&((\vec v_i -\vec v_j) \cdot \vec n_{ij} )\vec n_{ij}\ , \\
\vec v_{t_{ij}}&=&((\vec v_i -\vec v_j) - \vec v_{n_{ij}} - \frac{1}{2} (\sigma_i \vec \omega_i+\sigma_j \vec \omega_j) \times \vec n_{ij} \ .
\end{eqnarray}

The forces are modelled as~\cite{sun_forces, lee_forces}
\begin{eqnarray}
\vec F_{n_{ij}}&=&\sqrt{(\delta_{ij}/d)} (\kappa_n \delta_{n_{ij}} \vec n_{ij} -\gamma_n m_{eff} \vec v_{n_{ij}})\ ,\\
\vec F_{t_{ij}}&=&\sqrt{(\delta_{ij}/d)} (-\kappa_t \delta_{t_{ij}} \vec t_{ij} -\gamma_t m_{eff} \vec v_{t_{ij}}) \ ,
\label{forces}
\end{eqnarray}
with the  displacements  $\delta_{n_{ij}}= (\sigma_i+\sigma_j)/2-(\vec  r_i-\vec  r_j)$ and  $\delta_{t_{ij}} \vec t_{ij}= \vec v_{t_{ij}} dt$ in the normal and tangential directions, respectively.
The parameters $\kappa_n$ and $\kappa_t$ are the stiffness coefficients in the normal and tangential direction, respectively. The energy dissipated during the duration of the contact is regulated by the damping coefficients $\gamma_n$ and $\gamma_t$ and the effective mass $m_{eff}=m/2$. 
In addition, we model the static friction  by keeping track of the elastic shear displacement $\delta_{t_{ij}}$ over the contact lifetime and truncate it such that the condition $| F_{t_{ij}} |< |\mu F_{n_{ij}}|$ is satisfied, where  $\mu$ is the static friction coefficient. 

The force $\vec F_{iw}$ controlling the collision between particle $i$ and the container's  wall  are modelled by Eqs. (\ref{forces}) with the wall being described by a particle  with infinite mass leading to an effective mass $m_{eff}=m$.

The damping force between the particles and the bottom of the container is $ F_{b}~=~-\gamma_{b} m (\vec v_{i} -\vec v_{\rm bot})$, where
 $\gamma_{b}$ is a damping coefficient and $\vec v_{i}$ and $\vec v_{\rm bot}  $ are  the velocities of particle $i$ and of the bottom wall, respectively.

Once the forces on all particles are known the total  force $\vec F_i$ and torque $\vec \tau_i$ on a particle $i$ is determined by
\begin{eqnarray}
\vec F_i&=&\sum_j (\vec F_{n_{ij}}  +\vec F_{t_{ij}} ) -\gamma_{b} m (\vec v_{i} -\vec v_{\rm bot}) \ ,\\ \nonumber 
\vec \tau_i&=&-\frac{1}{2} \sum_j \sigma_j \vec n_{ij} \times \vec F_{t_{ij}} \ .
\end{eqnarray}

\begin{figure}[htbp]\centering
\includegraphics[width=8cm]{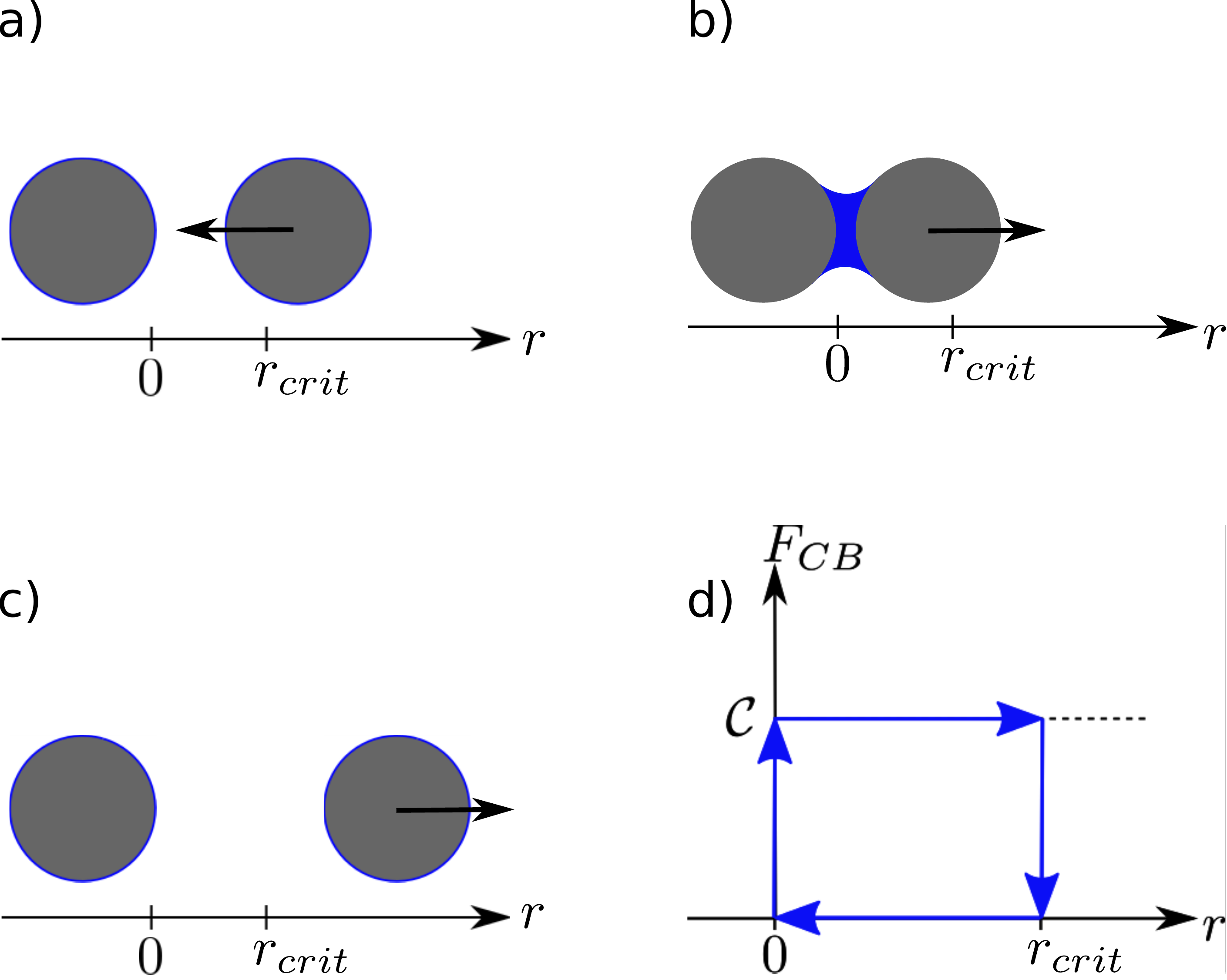}
\caption{(Colour online) Hysteretic formation and dissolution process of a capillary bridge between two particles. Arrows indicate the direction of particle motion. a) The distance between the particles is smaller than $r_{crit}$, but no capillary bridge is formed, because no previous contact between two particles occurred. b) After the collision a bridge is formed. c) The bridge dissolves as the distance between the particles becomes larger than $r_{crit}$. d) Force diagram for the capillary bridge interaction.}
\label{cap_bridge}
\end{figure}

Wet granular particles interact via additional forces caused by the formation and dissolution of capillary bridges.  This complex process is modelled in a simple way by the minimal capillary model (MCM)~\cite{herminghaus} that describes the bonding due to capillary bridges via an hysteretic cycle  (Fig.~\ref{cap_bridge}). The bridge between a pair of particles is only formed upon first contact, i.e. the interaction is zero until contact occurs. Subsequently the particles experience a constant attractive force $|\vec F_{CB}| = \mathcal C$ until a critical distance, $r_{crit}$ is reached. For separation distances larger than the critical distance the force is zero. We do not consider capillary bridges formation between the particles and the wall of the container.

The fundamental units of our simulation model are the particle mass $m$, the particle diameter  $\sigma$ and the gravitational acceleration $g$. Consequently, the derived units are the time $t_0=\sqrt{\sigma/g}$, velocity $v_0=\sqrt{g \sigma}$, force $f_0=m g$, elastic constant $k_0=m g/\sigma$, and damping coefficient $\gamma_{0}= \sqrt{g/\sigma}$.
The numerical values of the simulation parameters  are shown in Table I. 

\subsection{Simulation details}

We carry out  computer simulations using a standard Molecular Dynamics technique, in which the equations of motion are integrated via a Velocity-Verlet algorithm~\cite{Allen1987,Frenkel2002}.
The time step of the simulation is $\delta t \approx t_c / 50$~\cite{lee_forces}, where the contact time $t_c$ is estimated by~\cite{schaefer}
\begin{equation}
t_c = \pi \left( \frac{k_n}{m_{eff}}-\frac{\gamma_n^2}{4} \right)^{-0.5} \ .
\label{t_c}
\end{equation}

\begin{table}[htbp]
\caption{Parameters used in the simulation.}
\begin{center}
\begin{tabular}{cc}
\hline
Particles $k_n$ & $10^6~(k_{0})$\\
Particles $k_t$ & $\frac{2}{7} k_n$\\
Particles $\mu$ & 0.9\\
Particles $\gamma_{n}$ & 60 ($\gamma_{0})$\\
Particles $\gamma_{t}$ & 0  ($\gamma_{0})$\\
Wall $k_n$ & $10^6~(k_{0})$\\
Wall $k_t$ & $\frac{2}{7} k_n$\\
Wall $\mu$ & 0.5\\
Wall $\gamma_{n}$ & 60 ($\gamma_{0})$\\
Wall $\gamma_{t}$ & 30  ($\gamma_{0})$\\
\hline
$\mathcal C$ & 10~($f_{0})$\\
$r_{crit}$ & 0.1~($\sigma$)\\
\hline
\end{tabular}
\end{center}
\label{tab}

\end{table}%

The circular container  lies in the $x$-$y$ plane and is driven in a swirling motion according to 
\begin{eqnarray}
x_{c}&=&A \sin{( 2 \pi/P t)} \nonumber  \ ,\\
y_{c}&=&A \cos{ (2 \pi/P t)} \nonumber  \ ,
\end{eqnarray} 
where $x_{c}$ and $y_{c}$ are the coordinates of the centre of the  container, $ A $ is the amplitude of the oscillation, $P$ is the period of the oscillation and $ t $ is the time.

The particles in the container are initialised at random and the simulation  is carried out for $n=100$ swirling cycles.
During the simulation we analyse the local structure of each particle by means of the $q_{6}$ local bond order parameter, which allows us to distinguish between particles in a fluid or crystalline environment. Furthermore, we distinguish between hexagonal and square symmetries for the crystalline particles. 
Details about the $q_{6}$ implementation are given in appendix~\ref{sec:bondorder}.

\section{Results}
\label{res}
\subsection{Order-Disorder transition  for dry granular disks }

\begin{figure}[htbp]\centering
\includegraphics[width=8cm]{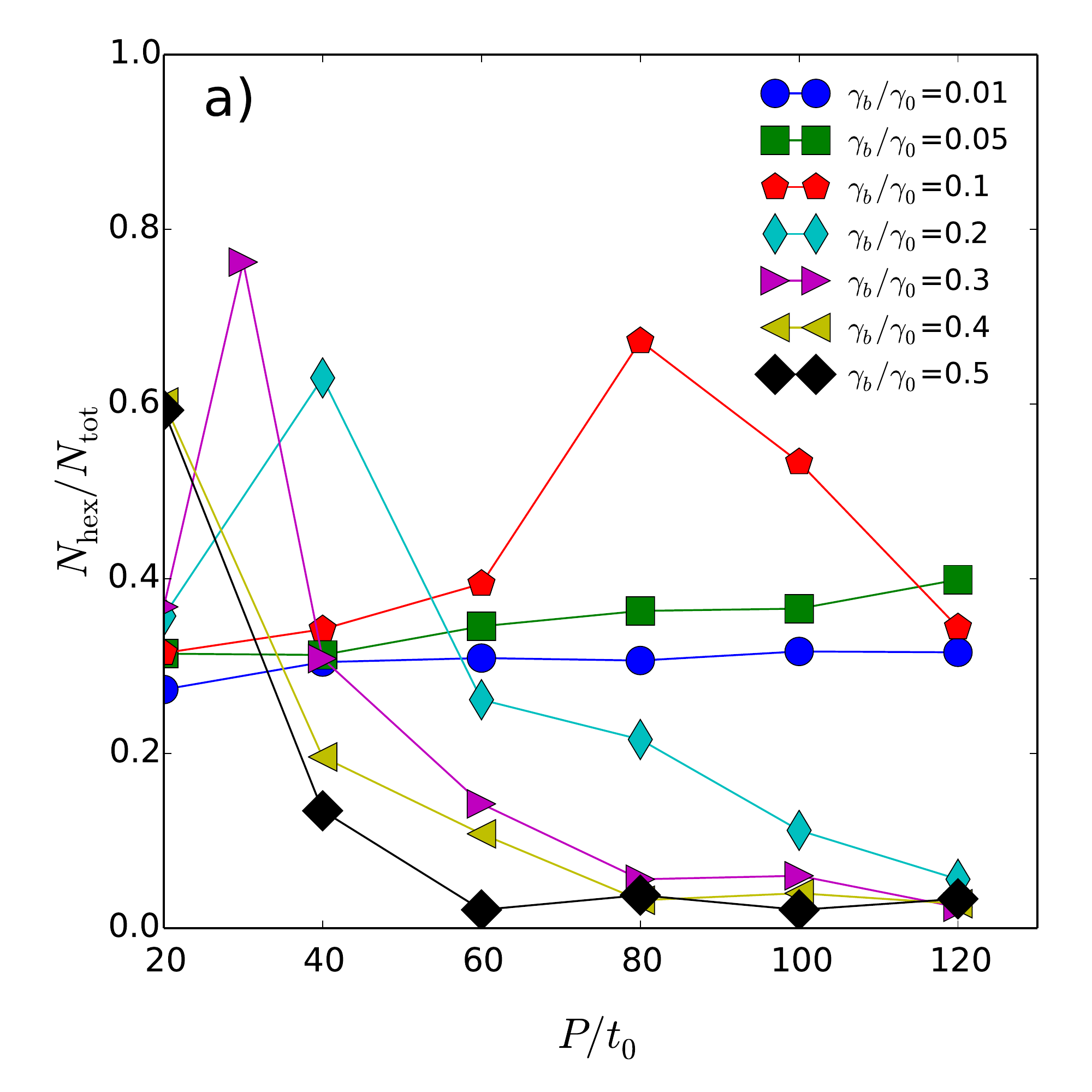}
\includegraphics[width=8cm]{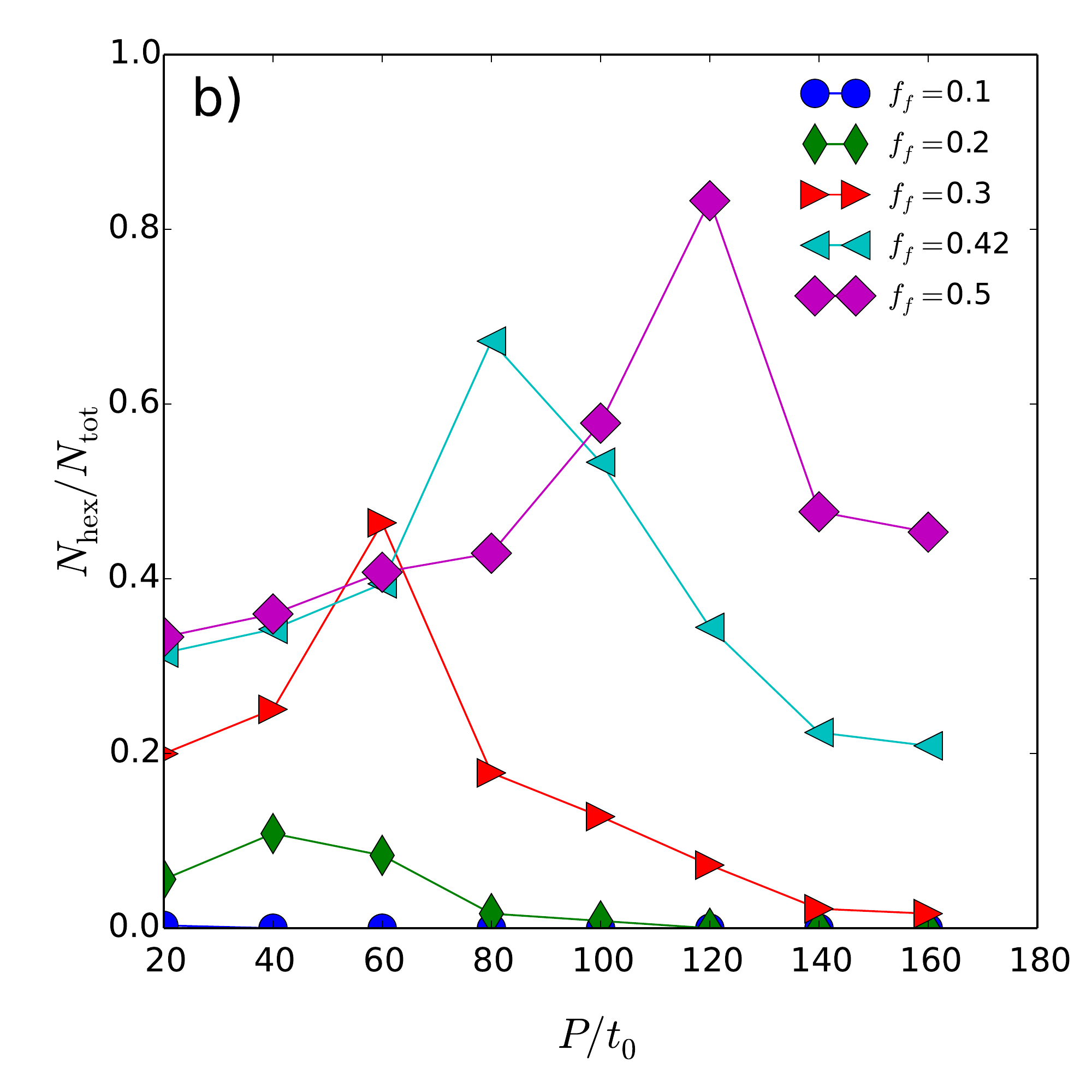}
\caption{(Colour online) Fraction of dry disks with hexagonal order  as a function of the oscillation period $P/t_{0}$. a) For a fixed filling fraction $f_{f}$=0.42 and different values of the bottom wall damping coefficient $\gamma_{b}/\gamma_{0}$. b) For a fixed $\gamma_{b}/\gamma_{0}=0.1$ and different filling fractions $f_{f}$. }
\label{damp}
\end{figure}

We begin the investigation of the dry granular disks by evaluating the effect of the energy dissipation that occurs due to the contact of the disks with the bottom wall. 
To this end we fix the amplitude of oscillation $A/\sigma=6$ and the number of particles $N=250$.
The fraction of particles with hexagonal  order $f_{\rm hex}=N_{\rm hex}/N_{\rm tot}$ is plotted against the periods of oscillation $P/t_{0}$ for different values of the bottom damping coefficient $\gamma_{b}/\gamma_{0}$ in Fig.~\ref{damp}a).

For $\gamma_{b}/\gamma_{0}=0.01$ and 0.05 we find an almost constant value of $f_{\rm hex} \simeq 0.3$ for all periods of oscillation while at $\gamma_{b}/\gamma_{0}=0.1$ we find a maximum $f_{\rm hex} \simeq 0.3$ at $P/t_{0}$=80. 
For increasing $\gamma_{b}$ (more dissipation) the location of the maximum $f_{\rm hex}$  moves toward decreasing values of the period $P$ (more input energy).

In Fig.~\ref{damp}b) we report the values of $f_{\rm hex}$ at different container filling fractions $f_{f}=N*\pi (\sigma/2)/\pi R^{2}$ and at a fixed damping coefficient $\gamma_{b}/\gamma_{0}=0.1$. 
At $N=250$ the filling fraction is $f_{f}=0.42$  the maximum of  $f_{\rm hex}$ at $P/t_{0}$=80 is recovered. 
At larger filling fractions we obtain maximum hexagonal order at larger periods of oscillation (smaller driving energy), while at smaller filling fractions, the maximum in  $f_{\rm hex}$ is obtained at smaller values of $P$. 
These results can be simply summarised by stating  that order is more easily achieved at high filling fractions.

\begin{figure}[htbf]\centering
\includegraphics[width=8cm]{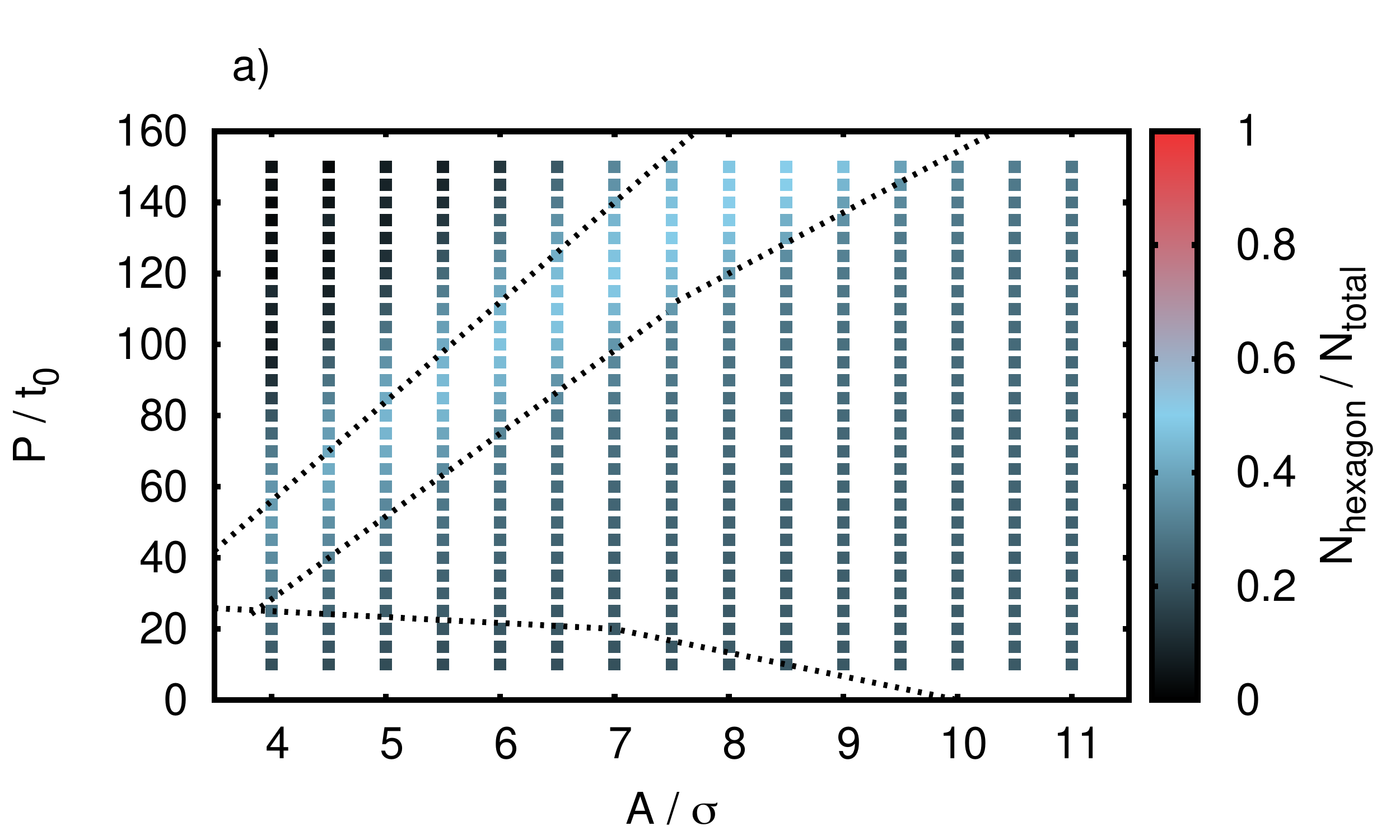}
\includegraphics[width=8cm]{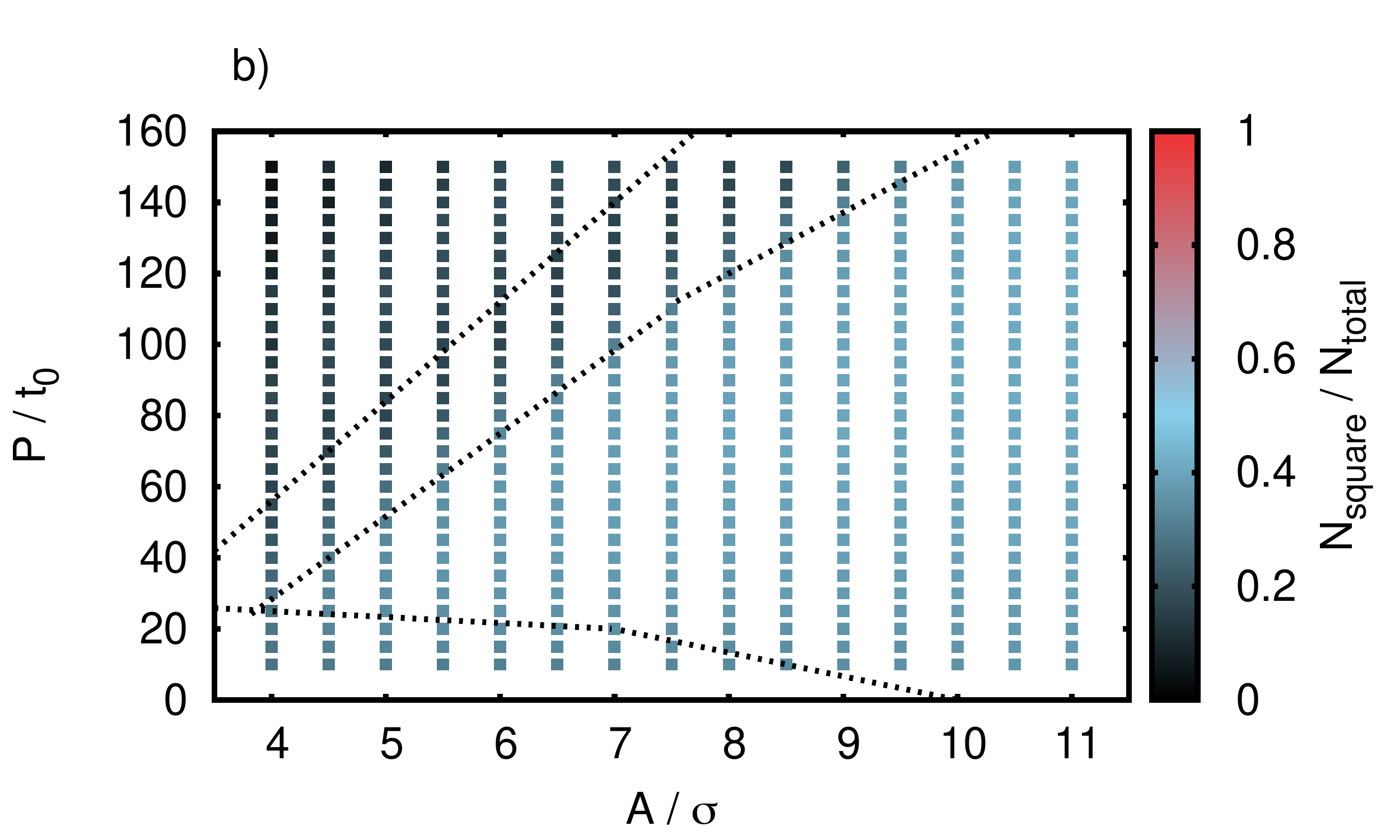}
\caption{(Colour online)  a) Fraction of dry disks with hexagonal local order. b) Fraction of particles with square local order. The dashed lines are guides to the eye to separate different phase states and crystal states.}
\label{dry_sq_hx}
\end{figure}

We now proceed with the systematic investigation of the order transition in dry granular disks at fixed $N=250$ (filling fraction $f_{f}$=0.42) and damping coefficient with the bottom wall $\gamma_{b}/\gamma_{0}=0.1$.
For each state point $(P/t_{0},A/\sigma)$ a computer simulation was run for 100 oscillations. We calculated the fraction of particles with hexagonal and square order using the $q_{6}$ local bond order parameter and  averaging its value over 80 oscillations. The results are summarised in Fig.~\ref{dry_sq_hx}. The fraction of particles with either hexagonal or square order is given by the colour of the points. The dashed lines are guides to the eye to distinguish regions with different predominant local order, and delimits region with a fraction of ordered particles larger than 0.25.

We find that at large periods of oscillation $P$ and small amplitudes $A$, corresponding to a low amount of driving energy,  most particles are in a disordered state, i.e  the fraction of square-ordered and hexagonal-ordered particles is less than 0.25. 
In this region of the diagram, the particles in the centre of the container are not affected by the wall due to the small oscillation amplitudes, while  the particles initially close to the wall are gently pushed toward the centre. 

We find a second region in the diagram that is rich in disordered particles, namely at small amplitudes and small periods of oscillation. Here the particles close to the wall are squeezed strongly to the wall while the particle in the centre remain basically unaffected by the container's motion because of the small amplitudes. 

At intermediate  driving energies the phase diagram is dominated by hexagonal order. Interestingly, an increase in  driving energy leads to a decrease of hexagonally ordered particles and a dramatic increase in particles with local  square order. 

It is worth to stress that in dry granular disks no attractive interaction between the disks is present. The formation of hexagonal order might be related to the entropically driven order transition of purely repulsive  colloidal systems.
On the other hand the square rich regions does not have a counterpart in  equilibrium colloidal systems. 

\begin{figure}[htbp]\centering
	\subfigure[\ ]{\includegraphics[width=4cm]{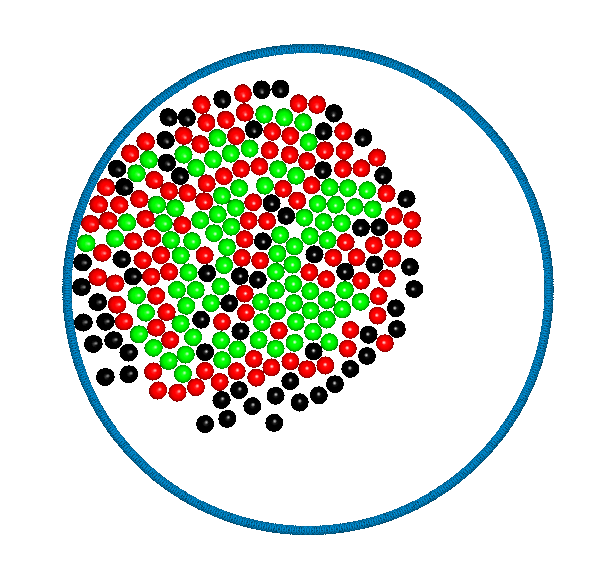}}
	\subfigure[\ ]{\includegraphics[width=4cm]{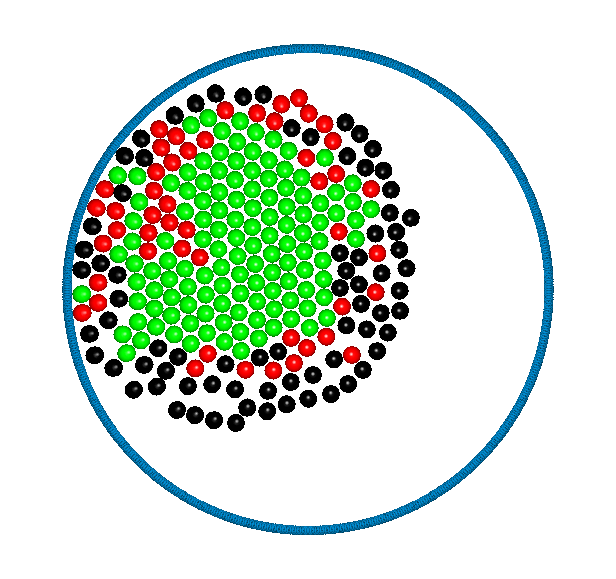}}
	\subfigure[\ ]{\includegraphics[width=4cm]{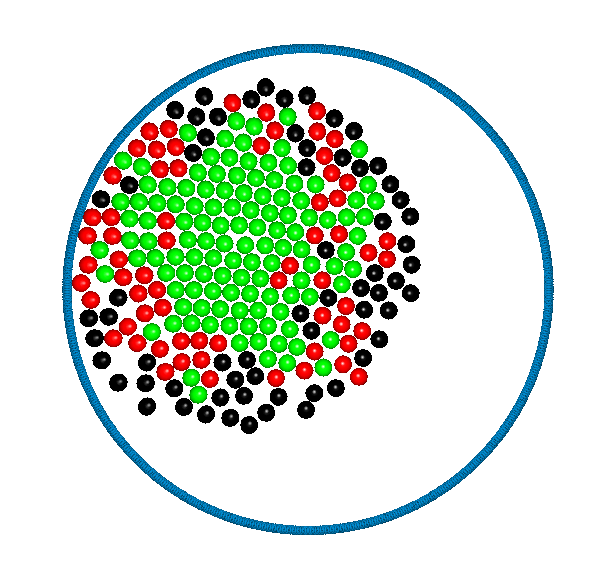}}
	\subfigure[\ ]{\includegraphics[width=4cm]{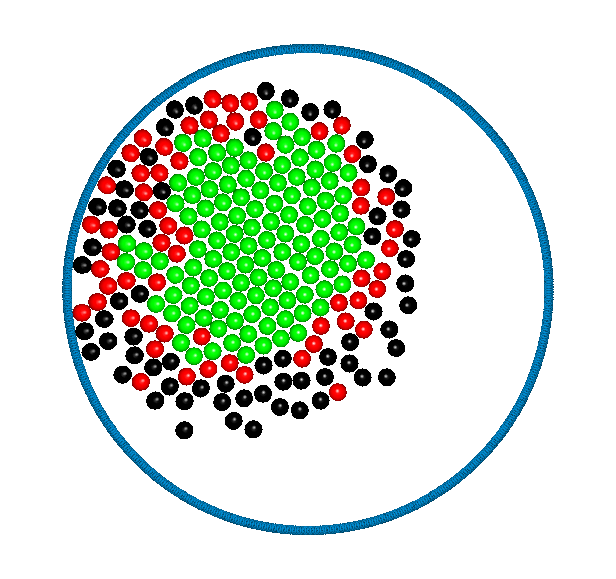}}
\caption{(Colour online) Simulation snapshots of dry disks at amplitude $A/\sigma = 7$ and period $P/t_{0}=120$. The green (light grey) disks have hexagonal order, the red (dark grey) disks have square order and the black particles are disordered.  a) $t/P=1$. b) $t/P=5. $  c)   $t/P=25. $  d) $t/P=100$.}
\label{snaps_A7}
\end{figure}

More insights in the order-disorder transition can be gained by looking at simulation snapshots.
The snapshots in Fig.~\ref{snaps_A7} are taken during a simulation with amplitude  $A/\sigma = 7$ and  period  $P/t_{0}=120$. In all snapshots the green disks have hexagonal order, the red disk have square order and the blue particles are disordered. After 1 oscillation (Fig.~\ref{snaps_A7} a)  the particles are mainly disordered and few clusters of hexagonal order are visible. After 5 oscillations (Fig.~\ref{snaps_A7} b) )we find a large cluster with circular shape. Particles with hexagonal local order are in the centre of the container. Around it we find square ordered and disordered particles. The overall structure remains qualitatively unchanged at later times  (Fig.~\ref{snaps_A7} c-d) . 

\begin{figure}[htbp]\centering
	\subfigure[\ ]{\includegraphics[width=4cm]{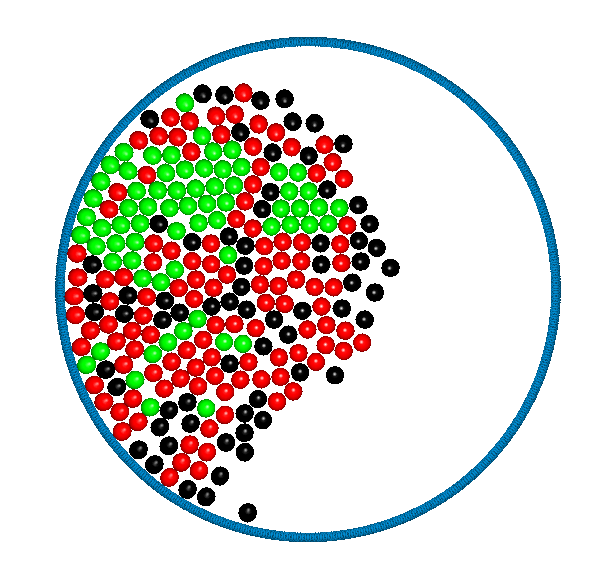}}
	\subfigure[\ ]{\includegraphics[width=4cm]{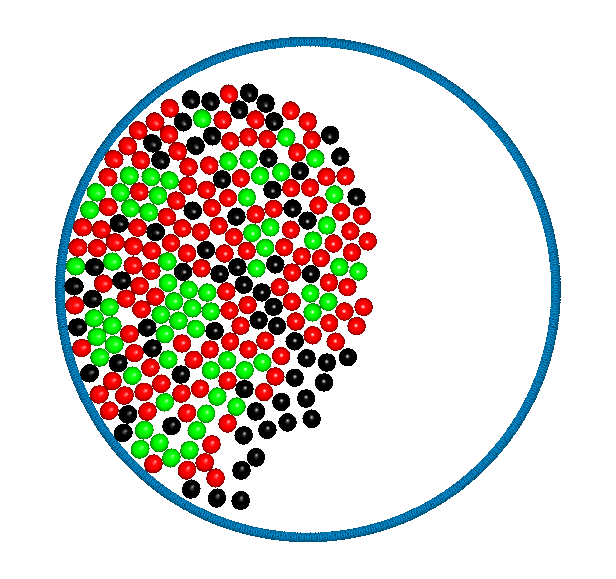}}
	\subfigure[\ ]{\includegraphics[width=4cm]{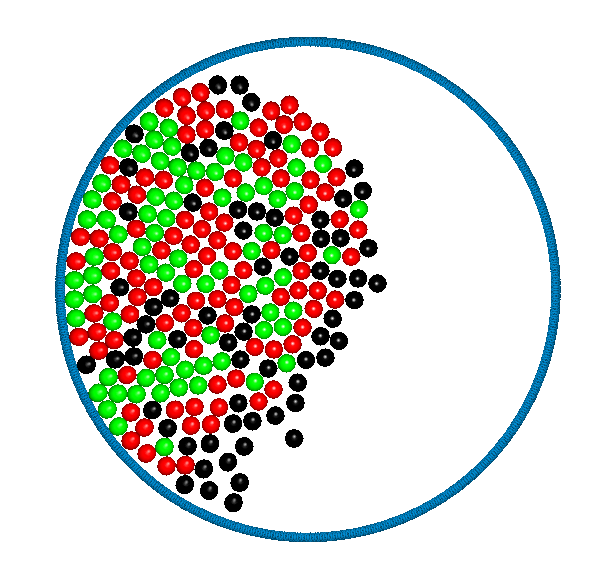}}
	\subfigure[\ ]{\includegraphics[width=4cm]{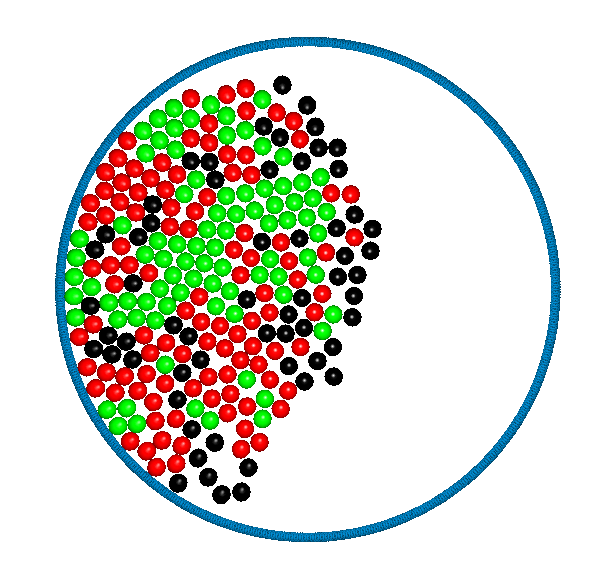}}
\caption{(Colour online) Simulation snapshots of dry disks  at amplitude $A/\sigma = 10$ and period $P/t_{0}=90$. The green (light grey) disks have hexagonal order, the red (dark grey) disks have square order and the black particles are disordered.  a) $t/P=1$. b) $t/P=5. $  c)   $t/P=25. $  d) $t/P=100$.}
\label{snaps_A10}
\end{figure}

By increasing the external driving energy we see how the structure changes dramatically (Fig.~\ref{snaps_A10} at amplitude $A/\sigma = 10$ and period $P/t_{0}=90$). We note that the granular cluster is slightly deformed and the particles show a large amount of square local order.

\subsection{Order-Disorder transition  for wet granular disks}

\begin{figure}[htbf]\centering
\includegraphics[width=8cm]{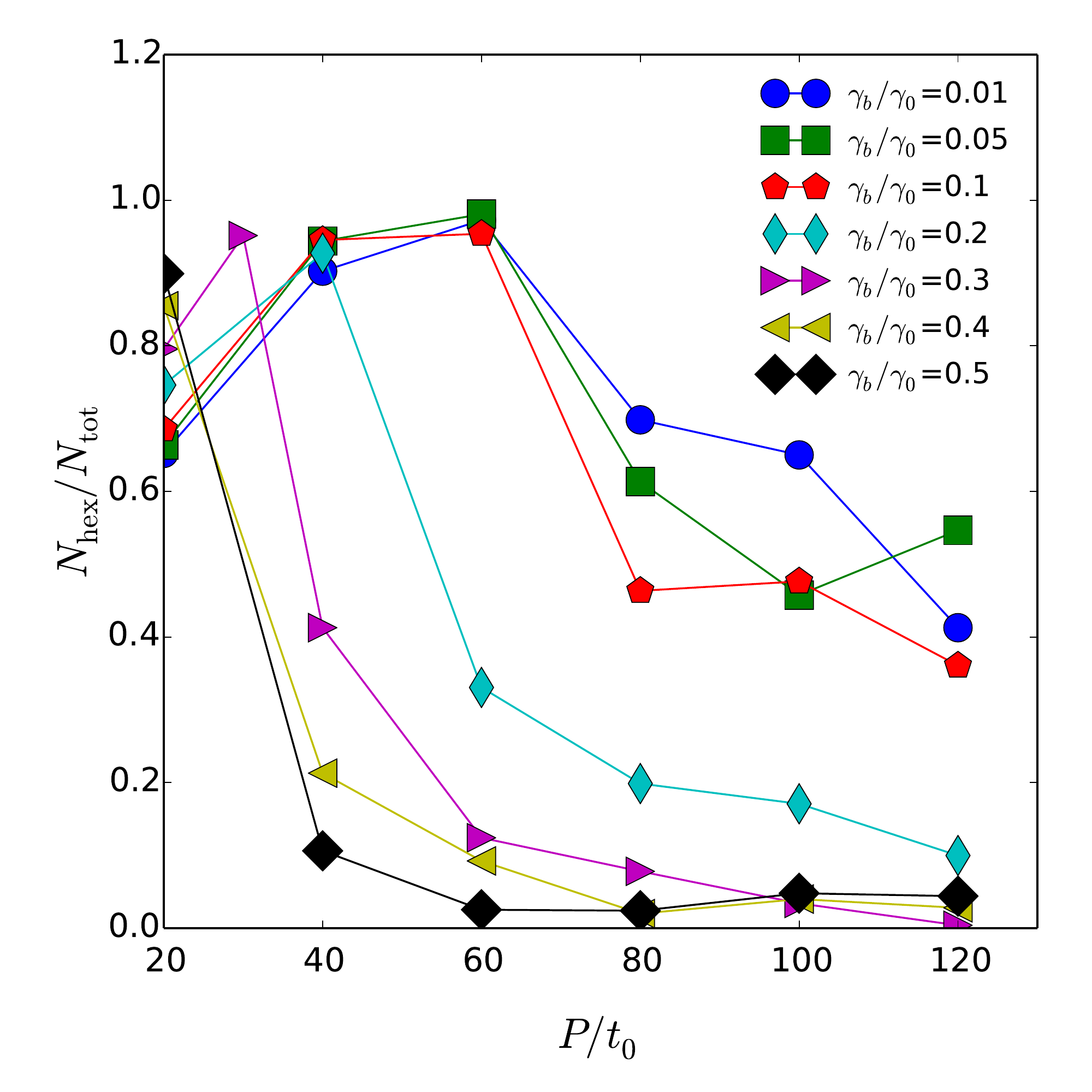}
\caption{(Colour online) Fraction of wet disks with hexagonal order  as a function of the oscillation period $P/t_{0}$,  for a fixed filling fraction $f_{f}$=0.42 and different values of the bottom wall damping coefficient $\gamma_{b}/\gamma_{0}$.}
\label{bottomwet}
\end{figure}

For wet granular particles the Fig.~\ref{bottomwet} shows the fraction of particles with hexagonal  order $f_{\rm hex}=N_{\rm hex}/N_{\rm tot}$ plotted against the period $P/t_{0}$ for different values of the damping coefficient $\gamma_{b}$ (dissipation at the bottom wall). 

In  contrast with the case of dry disks that showed a maximum fraction of hexagonally order particles of about 0.6, wet disks reach a fraction of hexagonally ordered particles of 1.
The dashed lines are guides to the eye to distinguish regions a fraction of ordered particles larger than 0.6.

For $\gamma_{b}/\gamma_{0} \leq 0.1$ there is almost no dependence on the bottom damping coefficient. The attractive interaction allows the formation of hexagonal ordered clusters even when there is little dissipation with the bottom wall.
For coefficients $\gamma_{b}/\gamma_{0} \geq 0.2$ the maximum order occurs at always smaller periods of oscillations, similar to the behaviour described for dry disks. 

\begin{figure}[htbf]\centering
\includegraphics[width=8cm]{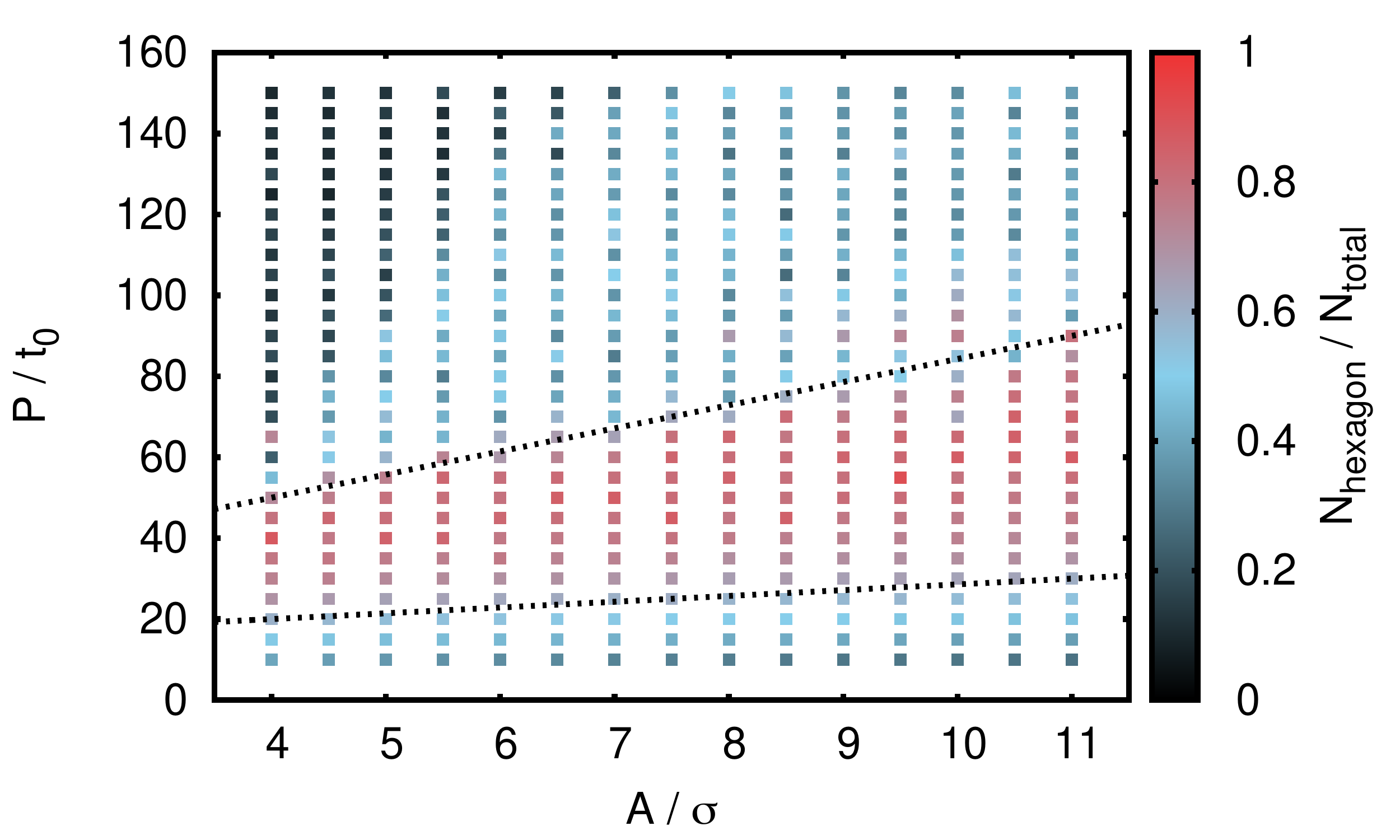}
\caption{(Colour online) Fraction of particles with hexagonal local order for  a system with $N$=250 wet granular disks. The dashed lines are guides to the eye.}
\label{wet_sq_hx}
\end{figure}

\begin{figure}[htbp]\centering
	\subfigure[\ ]{\includegraphics[width=5cm]{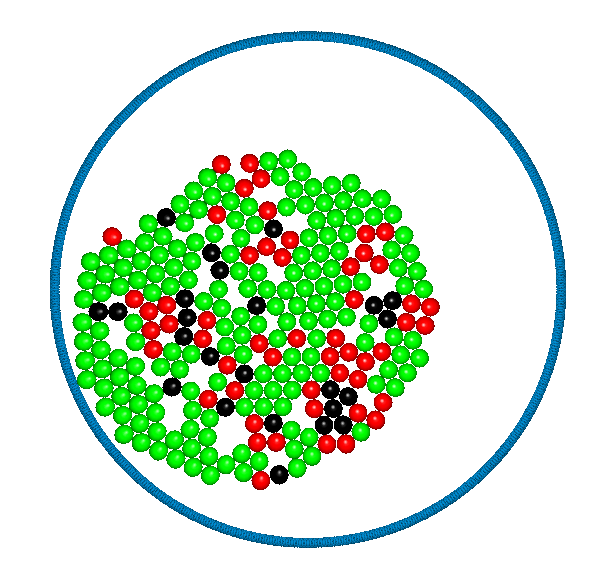}}
	\subfigure[\ ]{\includegraphics[width=5cm]{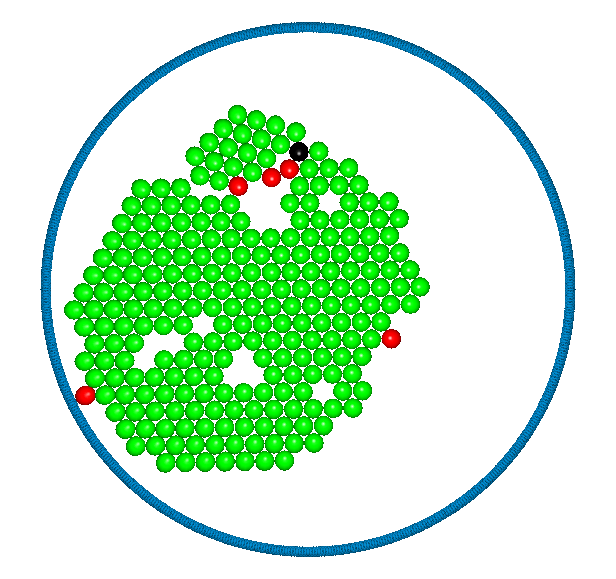}}
	\subfigure[\ ]{\includegraphics[width=5cm]{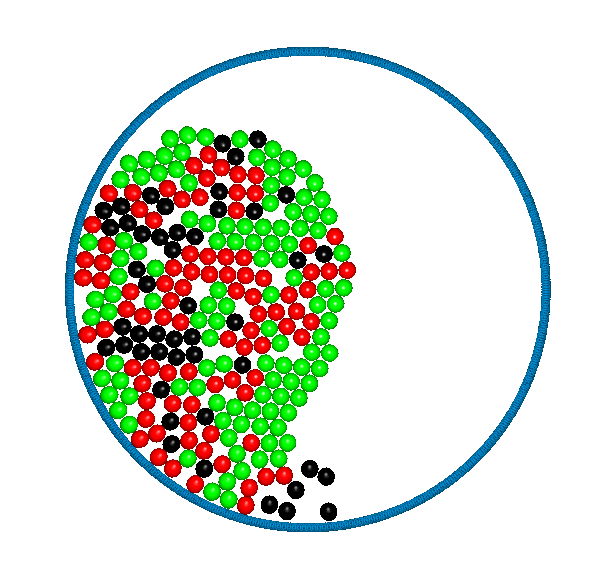}}
\caption{(Colour online) Simulation snapshots of wet disks at amplitude $A/\sigma = 9.5$ after 100 oscillations and different values of the oscillation period $P/t_{0}$. The green (light grey) disks have hexagonal order, the red (dark grey) disks have square order and the black particles are disordered. a) $P/t_{0}=95$. b) $P/t_{0}=55$.  c)   $P/t_{0}=15$.  }
\label{snaps_A95}
\end{figure}

We repeated the systematic study of the order-disorder transition for the wet particles. The results are summarised in Fig.~\ref{wet_sq_hx}. 
The region with predominant hexagonal order (Fig.~\ref{snaps_A95}b) is shifted, with respect the case of dry disks, toward smaller periods of oscillations and larger amplitudes, i.e. larger driving energy.
At larger energies the almost perfect order is destroyed and the particles are squeezed to the wall (Fig.~\ref{snaps_A95}c). At smaller energies, the particles form small clusters of hexagonally ordered clusters connected to each other via grain boundaries (Fig.~\ref{snaps_A95}a).

By comparing the diagram of Fig.~\ref{wet_sq_hx} for wet particles and the diagram of Fig.~\ref{dry_sq_hx} for dry disks we note that dry particles show a maximum order of 0.6 in a region where wet particles also have  an hexagonal order of about 0.5-0.6.  On the other hand, the structure of the overall granular cluster is very different in the two cases. Whereas for dry particles we find a large uniform cluster of hexagonally ordered particles (Fig.~\ref{snaps_A10}) , for wet particle the structure consist of small hexagonal clusters sparsely connected to each other (Fig.~\ref{snaps_A95}a).

\section{Conclusions}
\label{conc}
In conclusion, we investigated the order-disorder transition of dry and wet granular disks in a cylindrical container which is driven by a horizontal swirling motion.
We find that even for dry particles there is a well defined region of the parameter space, defined by the oscillation amplitude and period, where a large fraction of the particles show local hexagonal order.  
At larger driving energies the hexagonal order is destroyed and the particles are squeezed to the wall, leading to a large amount of particles with square local order. 

For the case of wet particles the attractive interaction of the capillary bridges enhances dramatically the amount of hexagonal order. Due to the strong attractive interaction between the disks mediated by the capillary bridges, the driving energy  required for the formation of circular clusters with high hexagonal order is higher for wet particles than for dry disks.

From the results of both dry and wet granular disk we can conclude that clusters of hexagonally ordered  particles occurs when a balance between the energy dissipation at the bottom wall (higher dissipation for higher $\gamma_{b}$)  external driving energy (higher energy for smaller $P$) is reached.

The dynamics  of the disordered structures was not investigated, but it would be interesting to see if  if the dynamical behaviour is related to the dynamics of the glass transition.
Furthermore, we point out that in our simulations of two dimensional disks we do not find the type of surface melting found in the experiments of~\citet{May:2013tf}. This result could be an indirect confirmation that the surface melting is due to the rolling friction of granular spheres with the bottom container as proposed in Ref.~\cite{May:2013tf} . We plan to carry out three dimensional simulations of our model system to confirm this  hypothesis and further investigate the origin and properties of the surface melting phenomenon.

The formation of compact ordered structures can influence  particle diffusion and consequently the segregation dynamics of differently sized particles. The understanding of the melting  and order formation phenomena can therefore lead to a better understanding of segregation phenomena. Further work is in progress to analyse   the dynamical properties of the phases reported in this article as well as the jamming transition and to relate them to the occurrence of segregation.

\appendix
\section{Bond order parameter} 
\label{sec:bondorder}

The local bond order parameter $q_l$ provides a measure of the local symmetry of a particle by looking at the distance and orientation of neighbouring particles. 
In general the bond order parameter is calculated according to~\cite{Steinhardt:1983hz}

\begin{equation}
	q_l(i)=\sqrt{ \frac{4 \pi}{2l+1} \sum_{m=-l}^{l} |q_{lm}(i)|^2} \quad,
	\label{ql}
\end{equation} 

where $q_{lm}(i)$ is defined as
\[ q_{lm}(i) = \frac{1}{N_b(i)} \sum_{j=1}^{N_b(i)} Y_{lm}(\varphi_{ij}, \vartheta_{ij}). \]
With $N_b(i)$ is the number of next neighbors of the particle $i$, $l$ is an integer, $m$ is an integer which runs from $-l$ to $l$ and $Y_{lm}$ are the spherical harmonics with the azimuthal angle between particle $i$ and $j$, $\varphi_{ij}$, and the polar angle, $\vartheta_{ij}$.\\
For a two-dimensional system we fix the angle $\vartheta_{ij} = 0$ for all pairs $i,j$.

Therefore, for the case $l=6$ the order parameter reads
\begin{equation}
	q_6(i)=\sqrt{ \frac{4 \pi}{13} \sum_{m=-6}^6 |q_{6m}(i)|^2} \quad ,
	\label{q6}
\end{equation}
with
\[ q_{6m} (i)= \frac{1}{N_b(i)} \sum_{j=1}^{N_b(i)} Y_{lm}(\varphi_{ij}, 0). \]
The angle $\varphi_{ij}$ is 
\[ \varphi_{ij} = \arctan \left( \frac{\vec r_{ij} \cdot \hat{\vec y}}{ \vec r_{ij} \cdot \hat{\vec x}} \right). \]
In two dimensions one can  distinguish three different types of order: chains, square and hexagonal. Table~\ref{tab:bondorder} gives the corresponding values of $q_6$.

\begin{table} [htbp] \centering 
\begin{tabular}{ccc} 
Structure & Coordination & $q_6$   \\ 
\hline 
Chain  &2& 1 \\
Square  &4 & 0.5863 \\
Hexagonal&6 & 0.7408 \\ 
\hline
\end{tabular}
\caption{Two dimensional structures that can be distinguished with the $q_{6}$ bond order parameter.}
\label{tab:bondorder}
\end{table}

\begin{acknowledgments}  
We thank Kai Huang, Matthias Schmidt and Ingo Rehberg for discussions. 
\end{acknowledgments}  
\bibliography{refs}

\begin{thebibliography}{23}%
\makeatletter
\providecommand \@ifxundefined [1]{%
 \@ifx{#1\undefined}
}%
\providecommand \@ifnum [1]{%
 \ifnum #1\expandafter \@firstoftwo
 \else \expandafter \@secondoftwo
 \fi
}%
\providecommand \@ifx [1]{%
 \ifx #1\expandafter \@firstoftwo
 \else \expandafter \@secondoftwo
 \fi
}%
\providecommand \natexlab [1]{#1}%
\providecommand \enquote  [1]{``#1''}%
\providecommand \bibnamefont  [1]{#1}%
\providecommand \bibfnamefont [1]{#1}%
\providecommand \citenamefont [1]{#1}%
\providecommand \href@noop [0]{\@secondoftwo}%
\providecommand \href [0]{\begingroup \@sanitize@url \@href}%
\providecommand \@href[1]{\@@startlink{#1}\@@href}%
\providecommand \@@href[1]{\endgroup#1\@@endlink}%
\providecommand \@sanitize@url [0]{\catcode `\\12\catcode `\$12\catcode
  `\&12\catcode `\#12\catcode `\^12\catcode `\_12\catcode `\%12\relax}%
\providecommand \@@startlink[1]{}%
\providecommand \@@endlink[0]{}%
\providecommand \url  [0]{\begingroup\@sanitize@url \@url }%
\providecommand \@url [1]{\endgroup\@href {#1}{\urlprefix }}%
\providecommand \urlprefix  [0]{URL }%
\providecommand \Eprint [0]{\href }%
\providecommand \doibase [0]{http://dx.doi.org/}%
\providecommand \selectlanguage [0]{\@gobble}%
\providecommand \bibinfo  [0]{\@secondoftwo}%
\providecommand \bibfield  [0]{\@secondoftwo}%
\providecommand \translation [1]{[#1]}%
\providecommand \BibitemOpen [0]{}%
\providecommand \bibitemStop [0]{}%
\providecommand \bibitemNoStop [0]{.\EOS\space}%
\providecommand \EOS [0]{\spacefactor3000\relax}%
\providecommand \BibitemShut  [1]{\csname bibitem#1\endcsname}%
\let\auto@bib@innerbib\@empty
\bibitem [{\citenamefont {Rosato}\ \emph {et~al.}(1987)\citenamefont {Rosato},
  \citenamefont {Strandburg}, \citenamefont {Prinz},\ and\ \citenamefont
  {Swendsen}}]{Rosato:1987dv}%
  \BibitemOpen
  \bibfield  {author} {\bibinfo {author} {\bibfnamefont {A.}~\bibnamefont
  {Rosato}}, \bibinfo {author} {\bibfnamefont {K.~J.}\ \bibnamefont
  {Strandburg}}, \bibinfo {author} {\bibfnamefont {F.}~\bibnamefont {Prinz}}, \
  and\ \bibinfo {author} {\bibfnamefont {R.~H.}\ \bibnamefont {Swendsen}},\
  }\href@noop {} {\bibfield  {journal} {\bibinfo  {journal} {Phys. Rev. Lett.}\
  }\textbf {\bibinfo {volume} {58}},\ \bibinfo {pages} {1038} (\bibinfo {year}
  {1987})}\BibitemShut {NoStop}%
\bibitem [{\citenamefont {Knight}\ \emph {et~al.}(1993)\citenamefont {Knight},
  \citenamefont {Jaeger},\ and\ \citenamefont {Nagel}}]{Knight:1993bg}%
  \BibitemOpen
  \bibfield  {author} {\bibinfo {author} {\bibfnamefont {J.~B.}\ \bibnamefont
  {Knight}}, \bibinfo {author} {\bibfnamefont {H.~M.}\ \bibnamefont {Jaeger}},
  \ and\ \bibinfo {author} {\bibfnamefont {S.~R.}\ \bibnamefont {Nagel}},\
  }\href@noop {} {\bibfield  {journal} {\bibinfo  {journal} {Phys. Rev. Lett.}\
  }\textbf {\bibinfo {volume} {70}},\ \bibinfo {pages} {3728} (\bibinfo {year}
  {1993})}\BibitemShut {NoStop}%
\bibitem [{\citenamefont {Cooke}\ \emph {et~al.}(1996)\citenamefont {Cooke},
  \citenamefont {Warr}, \citenamefont {Huntley},\ and\ \citenamefont
  {Ball}}]{Cooke:1996gg}%
  \BibitemOpen
  \bibfield  {author} {\bibinfo {author} {\bibfnamefont {W.}~\bibnamefont
  {Cooke}}, \bibinfo {author} {\bibfnamefont {S.}~\bibnamefont {Warr}},
  \bibinfo {author} {\bibfnamefont {J.~M.}\ \bibnamefont {Huntley}}, \ and\
  \bibinfo {author} {\bibfnamefont {R.~C.}\ \bibnamefont {Ball}},\ }\href@noop
  {} {\bibfield  {journal} {\bibinfo  {journal} {Phys. Rev. E}\ }\textbf
  {\bibinfo {volume} {53}},\ \bibinfo {pages} {2812} (\bibinfo {year}
  {1996})}\BibitemShut {NoStop}%
\bibitem [{\citenamefont {P{\"o}schel}\ and\ \citenamefont
  {Herrmann}(1995)}]{Poschel}%
  \BibitemOpen
  \bibfield  {author} {\bibinfo {author} {\bibfnamefont {T.}~\bibnamefont
  {P{\"o}schel}}\ and\ \bibinfo {author} {\bibfnamefont {H.~J.}\ \bibnamefont
  {Herrmann}},\ }\href@noop {} {\bibfield  {journal} {\bibinfo  {journal}
  {Europhys. Lett.}\ }\textbf {\bibinfo {volume} {29}},\ \bibinfo {pages} {123}
  (\bibinfo {year} {1995})}\BibitemShut {NoStop}%
\bibitem [{\citenamefont {Kudrolli}(2004)}]{Kudrolli:2004kr}%
  \BibitemOpen
  \bibfield  {author} {\bibinfo {author} {\bibfnamefont {A.}~\bibnamefont
  {Kudrolli}},\ }\href@noop {} {\bibfield  {journal} {\bibinfo  {journal} {Rep.
  Prog. Phys.}\ }\textbf {\bibinfo {volume} {67}},\ \bibinfo {pages} {209}
  (\bibinfo {year} {2004})}\BibitemShut {NoStop}%
\bibitem [{\citenamefont {Schr{\"o}ter}\ \emph {et~al.}(2006)\citenamefont
  {Schr{\"o}ter}, \citenamefont {Ulrich}, \citenamefont {Kreft}, \citenamefont
  {Swift},\ and\ \citenamefont {Swinney}}]{schroter:2006cc}%
  \BibitemOpen
  \bibfield  {author} {\bibinfo {author} {\bibfnamefont {M.}~\bibnamefont
  {Schr{\"o}ter}}, \bibinfo {author} {\bibfnamefont {S.}~\bibnamefont
  {Ulrich}}, \bibinfo {author} {\bibfnamefont {J.}~\bibnamefont {Kreft}},
  \bibinfo {author} {\bibfnamefont {J.~B.}\ \bibnamefont {Swift}}, \ and\
  \bibinfo {author} {\bibfnamefont {H.~L.}\ \bibnamefont {Swinney}},\
  }\href@noop {} {\bibfield  {journal} {\bibinfo  {journal} {Phys. Rev. E}\
  }\textbf {\bibinfo {volume} {74}},\ \bibinfo {pages} {011307} (\bibinfo
  {year} {2006})}\BibitemShut {NoStop}%
\bibitem [{\citenamefont {Majid}\ and\ \citenamefont
  {Walzel}(2009)}]{Majid:2009jl}%
  \BibitemOpen
  \bibfield  {author} {\bibinfo {author} {\bibfnamefont {M.}~\bibnamefont
  {Majid}}\ and\ \bibinfo {author} {\bibfnamefont {P.}~\bibnamefont {Walzel}},\
  }\href@noop {} {\bibfield  {journal} {\bibinfo  {journal} {Powder
  Technology}\ }\textbf {\bibinfo {volume} {192}},\ \bibinfo {pages} {311}
  (\bibinfo {year} {2009})}\BibitemShut {NoStop}%
\bibitem [{\citenamefont {Lagal}(1992)}]{lagal}%
  \BibitemOpen
  \bibfield  {author} {\bibinfo {author} {\bibfnamefont {R.}~\bibnamefont
  {Lagal}},\ }\href@noop {} {\emph {\bibinfo {title} {Gold Panning is Easy}}}\
  (\bibinfo  {publisher} {Ram Publishing Company, Dallas, TX},\ \bibinfo {year}
  {1992})\BibitemShut {NoStop}%
\bibitem [{\citenamefont {Aumaitre}\ \emph {et~al.}(2001)\citenamefont
  {Aumaitre}, \citenamefont {Kruelle},\ and\ \citenamefont
  {Rehberg}}]{Aumaitre:2001fd}%
  \BibitemOpen
  \bibfield  {author} {\bibinfo {author} {\bibfnamefont {S.}~\bibnamefont
  {Aumaitre}}, \bibinfo {author} {\bibfnamefont {C.~A.}\ \bibnamefont
  {Kruelle}}, \ and\ \bibinfo {author} {\bibfnamefont {I.}~\bibnamefont
  {Rehberg}},\ }\href@noop {} {\bibfield  {journal} {\bibinfo  {journal} {Phys.
  Rev. E}\ }\textbf {\bibinfo {volume} {64}},\ \bibinfo {pages} {041305}
  (\bibinfo {year} {2001})}\BibitemShut {NoStop}%
\bibitem [{\citenamefont {Schnautz}\ \emph {et~al.}(2005)\citenamefont
  {Schnautz}, \citenamefont {Brito}, \citenamefont {Kruelle},\ and\
  \citenamefont {Rehberg}}]{Schnautz:2005bp}%
  \BibitemOpen
  \bibfield  {author} {\bibinfo {author} {\bibfnamefont {T.}~\bibnamefont
  {Schnautz}}, \bibinfo {author} {\bibfnamefont {R.}~\bibnamefont {Brito}},
  \bibinfo {author} {\bibfnamefont {C.~A.}\ \bibnamefont {Kruelle}}, \ and\
  \bibinfo {author} {\bibfnamefont {I.}~\bibnamefont {Rehberg}},\ }\href@noop
  {} {\bibfield  {journal} {\bibinfo  {journal} {Phy. Rev. Lett.}\ }\textbf
  {\bibinfo {volume} {95}},\ \bibinfo {pages} {028001} (\bibinfo {year}
  {2005})}\BibitemShut {NoStop}%
\bibitem [{\citenamefont {Chung}\ \emph {et~al.}(2008)\citenamefont {Chung},
  \citenamefont {Ju},\ and\ \citenamefont {Liaw}}]{Chung:2008jz}%
  \BibitemOpen
  \bibfield  {author} {\bibinfo {author} {\bibfnamefont {F.~F.}\ \bibnamefont
  {Chung}}, \bibinfo {author} {\bibfnamefont {C.-Y.}\ \bibnamefont {Ju}}, \
  and\ \bibinfo {author} {\bibfnamefont {S.-S.}\ \bibnamefont {Liaw}},\
  }\href@noop {} {\bibfield  {journal} {\bibinfo  {journal} {Phys. Rev. E}\
  }\textbf {\bibinfo {volume} {77}},\ \bibinfo {pages} {061304} (\bibinfo
  {year} {2008})}\BibitemShut {NoStop}%
\bibitem [{\citenamefont {Scherer}\ \emph {et~al.}(1996)\citenamefont
  {Scherer}, \citenamefont {Buchholtz}, \citenamefont {P{\"o}schel},\ and\
  \citenamefont {Rehberg}}]{Scherer:1996hl}%
  \BibitemOpen
  \bibfield  {author} {\bibinfo {author} {\bibfnamefont {M.~A.}\ \bibnamefont
  {Scherer}}, \bibinfo {author} {\bibfnamefont {V.}~\bibnamefont {Buchholtz}},
  \bibinfo {author} {\bibfnamefont {T.}~\bibnamefont {P{\"o}schel}}, \ and\
  \bibinfo {author} {\bibfnamefont {I.}~\bibnamefont {Rehberg}},\ }\href@noop
  {} {\bibfield  {journal} {\bibinfo  {journal} {Phys. Rev. E}\ }\textbf
  {\bibinfo {volume} {54}},\ \bibinfo {pages} {R4560} (\bibinfo {year}
  {1996})}\BibitemShut {NoStop}%
\bibitem [{\citenamefont {Feltrup}\ \emph {et~al.}(2009)\citenamefont
  {Feltrup}, \citenamefont {Huang}, \citenamefont {Kruelle},\ and\
  \citenamefont {Rehberg}}]{Feltrup:2009gm}%
  \BibitemOpen
  \bibfield  {author} {\bibinfo {author} {\bibfnamefont {A.}~\bibnamefont
  {Feltrup}}, \bibinfo {author} {\bibfnamefont {K.}~\bibnamefont {Huang}},
  \bibinfo {author} {\bibfnamefont {C.~A.}\ \bibnamefont {Kruelle}}, \ and\
  \bibinfo {author} {\bibfnamefont {I.}~\bibnamefont {Rehberg}},\ }\href@noop
  {} {\bibfield  {journal} {\bibinfo  {journal} {Europ. Phys. J.:Special
  Topics}\ }\textbf {\bibinfo {volume} {179}},\ \bibinfo {pages} {19} (\bibinfo
  {year} {2009})}\BibitemShut {NoStop}%
\bibitem [{\citenamefont {K{\"o}tter}\ \emph {et~al.}(1999)\citenamefont
  {K{\"o}tter}, \citenamefont {Goles},\ and\ \citenamefont
  {Markus}}]{Kotter:1999iz}%
  \BibitemOpen
  \bibfield  {author} {\bibinfo {author} {\bibfnamefont {K.}~\bibnamefont
  {K{\"o}tter}}, \bibinfo {author} {\bibfnamefont {E.}~\bibnamefont {Goles}}, \
  and\ \bibinfo {author} {\bibfnamefont {M.}~\bibnamefont {Markus}},\
  }\href@noop {} {\bibfield  {journal} {\bibinfo  {journal} {Phys. Rev. E}\
  }\textbf {\bibinfo {volume} {60}},\ \bibinfo {pages} {7182} (\bibinfo {year}
  {1999})}\BibitemShut {NoStop}%
\bibitem [{\citenamefont {Scherer}\ \emph {et~al.}(2000)\citenamefont
  {Scherer}, \citenamefont {K{\"o}tter}, \citenamefont {Markus}, \citenamefont
  {Goles},\ and\ \citenamefont {Rehberg}}]{Scherer:2000jf}%
  \BibitemOpen
  \bibfield  {author} {\bibinfo {author} {\bibfnamefont {M.~A.}\ \bibnamefont
  {Scherer}}, \bibinfo {author} {\bibfnamefont {K.}~\bibnamefont {K{\"o}tter}},
  \bibinfo {author} {\bibfnamefont {M.}~\bibnamefont {Markus}}, \bibinfo
  {author} {\bibfnamefont {E.}~\bibnamefont {Goles}}, \ and\ \bibinfo {author}
  {\bibfnamefont {I.}~\bibnamefont {Rehberg}},\ }\href@noop {} {\bibfield
  {journal} {\bibinfo  {journal} {Phys. Rev. E}\ }\textbf {\bibinfo {volume}
  {61}},\ \bibinfo {pages} {4069} (\bibinfo {year} {2000})}\BibitemShut
  {NoStop}%
\bibitem [{\citenamefont {May}\ \emph {et~al.}(2013)\citenamefont {May},
  \citenamefont {Wild}, \citenamefont {Rehberg},\ and\ \citenamefont
  {Huang}}]{May:2013tf}%
  \BibitemOpen
  \bibfield  {author} {\bibinfo {author} {\bibfnamefont {C.}~\bibnamefont
  {May}}, \bibinfo {author} {\bibfnamefont {M.}~\bibnamefont {Wild}}, \bibinfo
  {author} {\bibfnamefont {I.}~\bibnamefont {Rehberg}}, \ and\ \bibinfo
  {author} {\bibfnamefont {K.}~\bibnamefont {Huang}},\ }\href@noop {}
  {\bibfield  {journal} {\bibinfo  {journal} {Phys. Rev. E}\ }\textbf {\bibinfo
  {volume} {88}},\ \bibinfo {pages} {062201} (\bibinfo {year}
  {2013})}\BibitemShut {NoStop}%
\bibitem [{\citenamefont {Sun}\ \emph {et~al.}(2006)\citenamefont {Sun},
  \citenamefont {Battaglia},\ and\ \citenamefont {Subramaniam}}]{sun_forces}%
  \BibitemOpen
  \bibfield  {author} {\bibinfo {author} {\bibfnamefont {J.}~\bibnamefont
  {Sun}}, \bibinfo {author} {\bibfnamefont {F.}~\bibnamefont {Battaglia}}, \
  and\ \bibinfo {author} {\bibfnamefont {S.}~\bibnamefont {Subramaniam}},\
  }\href@noop {} {\bibfield  {journal} {\bibinfo  {journal} {Phys. Rev. E}\
  }\textbf {\bibinfo {volume} {74}},\ \bibinfo {pages} {061307} (\bibinfo
  {year} {2006})}\BibitemShut {NoStop}%
\bibitem [{\citenamefont {Lee}(1994)}]{lee_forces}%
  \BibitemOpen
  \bibfield  {author} {\bibinfo {author} {\bibfnamefont {J.}~\bibnamefont
  {Lee}},\ }\href@noop {} {\bibfield  {journal} {\bibinfo  {journal} {J. Phys.
  A}\ }\textbf {\bibinfo {volume} {27}},\ \bibinfo {pages} {L257} (\bibinfo
  {year} {1994})}\BibitemShut {NoStop}%
\bibitem [{\citenamefont {Herminghaus}(2005)}]{herminghaus}%
  \BibitemOpen
  \bibfield  {author} {\bibinfo {author} {\bibfnamefont {S.}~\bibnamefont
  {Herminghaus}},\ }\href@noop {} {\bibfield  {journal} {\bibinfo  {journal}
  {Advances in Physics}\ }\textbf {\bibinfo {volume} {54}},\ \bibinfo {pages}
  {221} (\bibinfo {year} {2005})}\BibitemShut {NoStop}%
\bibitem [{\citenamefont {Allen}\ and\ \citenamefont
  {Tildesley}(1987)}]{Allen1987}%
  \BibitemOpen
  \bibfield  {author} {\bibinfo {author} {\bibfnamefont {M.~P.}\ \bibnamefont
  {Allen}}\ and\ \bibinfo {author} {\bibfnamefont {D.~J.}\ \bibnamefont
  {Tildesley}},\ }\href@noop {} {\emph {\bibinfo {title} {Computer Simulation
  of Liquids}}}\ (\bibinfo  {publisher} {Oxford University Press},\ \bibinfo
  {address} {New York},\ \bibinfo {year} {1987})\BibitemShut {NoStop}%
\bibitem [{\citenamefont {Frenkel}\ and\ \citenamefont
  {Smit}(2002)}]{Frenkel2002}%
  \BibitemOpen
  \bibfield  {author} {\bibinfo {author} {\bibfnamefont {D.}~\bibnamefont
  {Frenkel}}\ and\ \bibinfo {author} {\bibfnamefont {B.}~\bibnamefont {Smit}},\
  }\href@noop {} {\emph {\bibinfo {title} {Understanding {M}olecular
  {S}imulation 2nd edition}}},\ \bibinfo {series} {Computational Science
  Series}, Vol.~\bibinfo {volume} {1}\ (\bibinfo  {publisher} {Academic
  Press},\ \bibinfo {year} {2002})\BibitemShut {NoStop}%
\bibitem [{\citenamefont {Schaefer}\ \emph {et~al.}(1996)\citenamefont
  {Schaefer}, \citenamefont {Dippel},\ and\ \citenamefont {Wolf}}]{schaefer}%
  \BibitemOpen
  \bibfield  {author} {\bibinfo {author} {\bibfnamefont {J.}~\bibnamefont
  {Schaefer}}, \bibinfo {author} {\bibfnamefont {S.}~\bibnamefont {Dippel}}, \
  and\ \bibinfo {author} {\bibfnamefont {D.}~\bibnamefont {Wolf}},\ }\href@noop
  {} {\bibfield  {journal} {\bibinfo  {journal} {J. Phys. I France}\ }\textbf
  {\bibinfo {volume} {6}},\ \bibinfo {pages} {5} (\bibinfo {year}
  {1996})}\BibitemShut {NoStop}%
\bibitem [{\citenamefont {Steinhardt}\ \emph {et~al.}(1983)\citenamefont
  {Steinhardt}, \citenamefont {Nelson},\ and\ \citenamefont
  {Ronchetti}}]{Steinhardt:1983hz}%
  \BibitemOpen
  \bibfield  {author} {\bibinfo {author} {\bibfnamefont {P.}~\bibnamefont
  {Steinhardt}}, \bibinfo {author} {\bibfnamefont {D.}~\bibnamefont {Nelson}},
  \ and\ \bibinfo {author} {\bibfnamefont {M.}~\bibnamefont {Ronchetti}},\
  }\href@noop {} {\bibfield  {journal} {\bibinfo  {journal} {Physical Review
  B}\ }\textbf {\bibinfo {volume} {28}},\ \bibinfo {pages} {784} (\bibinfo
  {year} {1983})}\BibitemShut {NoStop}%
\end{thebibliography}%

\end{document}